\documentclass[amsmath,amssymb,12pt,aps]{revtex4-2}
\usepackage[utf8]{inputenc} 
\usepackage{graphicx}
\usepackage{subcaption}
\usepackage{float}
\usepackage{natbib}
\usepackage{wrapfig}
\usepackage{xcolor}
\usepackage{dcolumn}
\usepackage[english]{babel}
\newcommand{\floor}[1]{\left\lfloor #1 \right\rfloor}

\begin{document}

\title{Emergence of Collective Accuracy in Socially Connected Networks}

\author{Dan Braha$^{1,3,}$\footnote{Corresponding author}} 

\author{Marcus A. M. de Aguiar$^{1,2}$}

\affiliation{$^1$New England Complex Systems Institute, Cambridge, Massachusetts, United States of America}

\affiliation{$^2$Instituto de F\'isica `Gleb Wataghin', Universidade Estadual de Campinas, Unicamp 13083-970, Campinas, SP, Brazil}

\affiliation{$^3$University of Massachusetts, Dartmouth, Massachusetts, United States of America }

\begin{abstract}
	
We analyze the accuracy of collective decision-making in socially connected populations, where agents update binary choices through local interactions on a network. Each agent receives a private signal that is biased— even marginally—toward the correct alternative, and social influence mediates the aggregation of these signals. We show analytically that, in the large-population limit, the probability of a correct majority converges to a nontrivial expression involving the regularized incomplete beta function. Remarkably, this collective accuracy surpasses that of any individual agent whenever private signals are better than random—revealing that network-mediated influence can enhance, rather than impair, group performance. Our findings may inform the design of resilient decision-making systems in social, biological, and engineered networks, where accuracy must emerge from interdependent and noisy agents.
	
\end{abstract}

\maketitle

\newpage

\section{Introduction and motivation}

In this Letter, we show that collective decision-making in socially connected networks can achieve enhanced accuracy despite peer influence, extending classical models beyond independent voting. Our approach builds upon the classical voter model, a well-established probabilistic framework for studying interacting particle systems, cooperative phenomena, and collective dynamics \cite{r1,r2,r3,r4,r5}.

We consider a network represented by a graph $G=(V,E)$, where the set of nodes $V$ corresponds to voters and edges $E$ denote pairwise interactions. Each of the $n+\alpha+\beta$ nodes selects between two alternatives: “correct” (state 1) or “incorrect” (state 0). The network structure includes $\alpha$ voters fixed in state 1, $\beta$ voters fixed in state 0, and $n$ free voters whose states evolve through peer influence. Voter models with fixed agents, or “zealots,” have provided significant insights into the effects of external biases, studied using approaches ranging from mean-field approximations \cite{r6,r7,r8,r9} to heterogeneous networks \cite{r10,r11,r12,r13}.

The network dynamics proceed in discrete time: at each step, a randomly selected free voter adopts the state of a randomly chosen neighbor. In the long-time limit $t \rightarrow \infty$, the alternative supported by more than half of the free voters is selected as the consensus outcome.

Let X denote the number of free voters choosing the correct alternative. In prior work \cite{r11,r14,r15,r16}, it was established that for a complete graph—where every node is connected to every other—the stationary probability of observing $X=k$ free voters in state 1, independent of the initial configuration, is:
\begin{equation}
	P\left(X=k\right)=\frac{\left(\genfrac{}{}{0pt}{0}{\alpha +k-1}k\right)\left(\genfrac{}{}{0pt}{0}{n+\beta
			-k-1}{n-k}\right)}{\left(\genfrac{}{}{0pt}{0}{n+\alpha +\beta -1}n\right)}.
	\label{eq1}
\end{equation}

Here, $\alpha$ and $\beta$ serve as external influence parameters, generalizable to positive real values through the gamma function. Notably, when $\alpha=\beta=1$, this distribution becomes uniform, independent of $n$, a characteristic unique to fully connected networks.

If the free voters act independently and copy only the states of the $\alpha+\beta$ fixed voters, the stationary probability of observing $k$ free voters in state 1 follows a binomial distribution:
\begin{equation}
	P\left(X=k\right)=\left(\genfrac{}{}{0pt}{0}nk\right)p^k\left(1-p\right)^{n-k},
\end{equation}
where $p=\alpha⁄(\alpha+\beta)$ represents the probability that a free voter selects the correct alternative.

According to Condorcet's Jury Theorem \cite{r17,r18}, independent voters with $p > 1/2$ yield a probability of collective correctness 
satisfying:
\begin{equation}
	P_n(correct \; decision) = P(X \geq \floor{n/2} + 1) = \sum_{k=\floor{n/2} + 1}^{n}  \binom{n}{k} p^k (1-p)^{n-k} > p
\end{equation}
with $P_n\left(\mathit{correct\ decision}\right)\rightarrow 1$ as $n\rightarrow {\infty}$. Condorcet’s theorem thus 
demonstrates that a group’s collective accuracy surpasses individual accuracy provided that individual signals are better than random.

In contrast, when peer influence is incorporated via a fully connected network, the analogous inequality:
\begin{equation}
	P_n(correct \; decision) = P(X \geq \floor{n/2} + 1) = \sum_{k=\floor{n/2} + 1}^{n}  \frac{\binom{\alpha+k-1}{k} \binom{n+\beta-k-1}{n-k}}{\binom{n+\alpha+\beta-1}{n}} > p
\end{equation}
does not hold universally for all $n$, though numerical studies suggest it remains valid asymptotically for large $n$. This Letter 
rigorously analyzes the asymptotic regime $n \rightarrow \infty$, establishing a novel connection between the group’s probability of 
making a correct decision and an inequality involving the Beta distribution. We demonstrate that peer influence, far from degrading group 
performance, can amplify collective accuracy beyond that achievable by isolated individuals. These results show that even in the presence 
of strong social dependence, collective wisdom can persist and even strengthen in large networks.

\section{Results}

To analyze the asymptotic behavior as $n \rightarrow \infty$, we define the proportion $V=X⁄n$, representing the fraction of free voters who select the correct alternative in steady state. We can then establish the following result:\\

\noindent {\bf Theorem} -  Let  $X$ \ denote the number of free voters who select the correct alternative. For a fully connected network, in
equilibrium, the proportion  $V=X/n$ \ converges in distribution to a Beta distribution:
\begin{equation}
	V=\frac{X}{n}   \overset{d}{\longrightarrow}  \mathit{Beta}\left(\alpha ,\beta \right)
\end{equation}
where  $\mathit{Beta}\left(\alpha ,\beta \right)$ \ represents a Beta random variable with parameters  $\alpha$ 
and $\beta $. Thus, as  $n\rightarrow {\infty}$, the probability of making a correct decision becomes:
\begin{equation}
	\lim _{n\rightarrow {\infty}}P_n\left(\mathit{correct}\;\mathit{decision}\right)=P\left(V{\geq}1/2\right)
	\label{eq21}
\end{equation}

\noindent {\it Proof}  -  Assume  $\alpha >0$ \ and  $\beta >0$. To derive the continuous probability density function for  $V$, 
we consider the limit  $n\rightarrow {\infty}$ \ while keeping  $V=X/n$ fixed. Partition the interval  $\left(0,1\right)$ into  $n$ 
subintervals, each of length  $1/n$, and fix  $V=v=k/n{\in}\left(0,1\right)$. Then the probability distribution satisfies
$P\left(V=k/n\right) = P\left(X=k\right)$, which can be interpreted as the area of a rectangle centered at 
$v=k/n$, with base  $1/n$ and height  $f\left(v\right)$. Thus:
\begin{equation}
	P\left(V=\frac k n\right)=P\left(X=k\right)=\frac{f(v)} n,
\end{equation}
or equivalently, $f\left(v\right)=\mathit{nP}\left(X=k\right)$.

As  $n\rightarrow {\infty}$, the width of each rectangle tends to zero, and the height approaches a limit:

\begin{equation}
	\lim _{n\rightarrow {\infty}}f\left(v\right)= \lim _{n\rightarrow {\infty}}\mathit{nP}\left(X=k\right),
\end{equation}
which gives the continuous probability density of  $V$. By taking the limit where both  $k=\mathit{vn}$ \ and 
$n-k=n\left(1-v\right)$ \ tend to infinity for fixed  $v$, we can derive the limiting probability density function.

We next express $P(X=k)$, as given by Eq.(\ref{eq1}), using Gamma functions:
\begin{equation}
	P\left(X=k\right)=\frac{\Gamma (\alpha +k)\Gamma (n+\beta -k)\Gamma (n+1)\Gamma (\alpha +\beta )}{\Gamma (k+1)\Gamma
		(\alpha )\Gamma (n-k+1)\Gamma (\beta )\Gamma (n+\alpha +\beta )}.
\end{equation}
Applying Stirling's approximation 
\begin{equation}
	\frac{\Gamma (x+t)}{\Gamma (x+u)} \rightarrow x^{t-u}
	\label{eq3}
\end{equation}
as $x\rightarrow {\infty}$, to each Gamma function as $n\rightarrow {\infty}$, we simplify
\begin{eqnarray}
	\lim _{n\rightarrow {\infty}}  &\mathit{nP}\left(X=k\right) =\lim _{n\rightarrow {\infty}}n\frac{\Gamma
		\left(n+1\right)}{\Gamma \left(n+\alpha +\beta \right)} \cdot \frac{\Gamma \left(\alpha +\beta \right)}{\Gamma
		\left(\alpha \right)\Gamma \left(\beta \right)} \cdot \frac{\Gamma \left(k+\alpha \right)}{\Gamma
		\left(k+1\right)} \cdot \frac{\Gamma \left(n-k+\beta \right)}{\Gamma \left(n-k+1\right)} \nonumber \\
	&=\lim _{n\rightarrow {\infty}}\frac n{n+1}(n+1) \cdot \frac{\Gamma \left(n+1\right)n^{2-\left(\alpha +\beta
			\right)}}{\Gamma \left(n+\alpha +\beta \right)n^{2-\left(\alpha +\beta \right)}} \cdot \frac{\Gamma \left(\alpha
		+\beta \right)}{\Gamma \left(\alpha \right)\Gamma \left(\beta \right)} \nonumber \\
	& \cdot \frac{\Gamma \left(\mathit{vn}+\alpha \right)\left(\mathit{vn}\right)^{\alpha -1}}{\Gamma \left(\mathit{vn}+1\right)\left(\mathit{vn}\right)^{\alpha
			-1}} \cdot \frac{\Gamma \left(n\left(1-v\right)+\beta \right)\left(n\left(1-v\right)\right)^{\beta -1}}{\Gamma
		\left(n\left(1-v\right)+1\right)\left(n\left(1-v\right)\right)^{\beta -1}}  \nonumber \\
	&=\underset{n\rightarrow {\infty}}{\lim }\left(\frac n{n+1}\right) \cdot \left(\frac{\Gamma \left(n+2\right)}{\Gamma
		\left(n+\alpha +\beta \right)n^{2-\left(\alpha +\beta \right)}}\right)  \nonumber \\
	& \cdot \left(\frac{\Gamma
		\left(\mathit{vn}+\alpha \right)}{\Gamma \left(\mathit{vn}+1\right)\left(\mathit{vn}\right)^{\alpha
			-1}}\right)  \cdot  \left(\frac{\Gamma \left(n\left(1-v\right)+\beta \right)}{\Gamma
		\left(n\left(1-v\right)+1\right)\left(n\left(1-v\right)\right)^{\beta -1}}\right) \nonumber \\
	&\cdot \frac 1{B\left(\alpha ,\beta 	\right)} \cdot v^{\alpha -1} \cdot \left(1-v\right)^{\beta -1} \nonumber \\
	&= \frac 1{B\left(\alpha ,\beta \right)} \cdot v^{\alpha -1} \cdot \left(1-v\right)^{\beta -1},
\end{eqnarray}
where  $B\left(\alpha ,\beta \right)=\frac{\Gamma \left(\alpha \right)\Gamma \left(\beta \right)}{\Gamma \left(\alpha +\beta \right)}$ is 
the Beta function. In the third equality, we used the identity  $\Gamma \left(n+2\right)=\left(n+1\right)\Gamma \left(n+1\right)$ 
and Stirling's approximation in Eq.(\ref{eq3}) to simplify the terms. Thus, $V$ converges to the $Beta(\alpha,\beta)$ distribution.

Moreover, for any  $x{\in}\left(0,1\right)$ 
\begin{equation}
	P\left(V{\leq}x\right)=P\left(X{\leq}\mathit{nx}\right)=\sum_{k=0}^{\floor{nx}} P(X=k)\rightarrow \frac 1{B(\alpha ,\beta )}\int _0^xt^{\alpha
		-1}(1-t)^{\beta -1}\mathit{dt},
\end{equation}
which is the cumulative distribution function of the Beta distribution. Hence,  $V$ \ converges in distribution to 
$\mathit{Beta}\left(\alpha ,\beta \right)$, completing the proof.

It is important to note that for sufficiently large values of  $\alpha $ \ and  $\beta $\ (i.e.,  $\alpha ,\beta \gg 1$), the influence of 
the external “fixed” voters dominates the dynamics of the free voters [11]. In this regime, individual decisions become effectively 
independent, and the random variable 
$V \sim \mathit{Beta}\left(\alpha ,\beta \right)$ \ can be approximated by a normal distribution \cite{r19}:
\begin{equation}
	V \sim N\left(\mu =\frac{\alpha }{\alpha +\beta },\sigma ^2=\frac{\mathit{\alpha \beta }}{\left(\alpha +\beta
		\right)^2(\alpha +\beta +1)}\right)
\end{equation}
As  $\alpha $ \ and  $\beta $ \ increase, the Beta distribution becomes sharply concentrated around its mean. As a
result, the probability of a correct majority decision, as specified in Eq.\ref{eq21}, tends to 1 if  $\mu =\frac{\alpha
}{\alpha +\beta }>\frac 1 2$, recovering the classical Condorcet's Jury Theorem in the appropriate limit.

However, for general values of \  $\alpha $ \ and  $\beta $ \ where  $\alpha /(\alpha +\beta) > 1/2$,   $\lim_{n \to \infty} 
P_n(\text{correct decision})$  $ = P\left(V \geq \frac{1}{2}\right)$ \ does not 
necessarily converge to 1. Nonetheless, numerical simulations indicate that
\begin{equation}
	\lim _{n\rightarrow {\infty}}P_n\left(\mathit{correct} \; \mathit{decision}\right)=P\left(V{\geq}1/2\right)>\frac{\alpha
	}{\alpha +\beta }.
	\label{eq4}
\end{equation}
We establish this bound analytically below, showing that when individual signals are biased toward the correct 
choice—even marginally—the probability that the majority selects the correct decision exceeds that of any individual. This result 
highlights how even weakly informative signals, when coupled through social interactions, can lead to collective amplification. 
Such a mechanism echoes cooperative phenomena in physical systems, where global order emerges from locally interacting components. 

The inequality in Eq. \ref{eq4} relates closely to properties of the Beta cumulative distribution function $F(x; \alpha,\beta)$ and associated special functions \cite{r16}. Specifically, letting $V \sim Beta(\alpha,\beta)$, we define:
\begin{equation}
	F\left(x;\alpha ,\beta \right) = P(V \leq x ) = I_x\left(\alpha ,\beta \right)
\end{equation}
where  $I_x\left(\alpha ,\beta \right)$ \ is the regularized incomplete Beta function: 
\begin{equation}
	I_x\left(\alpha ,\beta \right) = \frac{1}{B(\alpha ,\beta )} \int _0^x t^{\alpha -1}(1-t)^{\beta -1}\mathit{dt} 
\end{equation}
Thus, the inequality in Eq.\ref{eq4} can be formalized as the following proposition:\\

\noindent {\bf Proposition} Let $\alpha, \beta > 0$ with $\alpha > \beta$. Then:
	\[
	P(V \geq \tfrac{1}{2}) = 1 - I_{1/2}(\alpha, \beta) > \frac{\alpha}{\alpha + \beta},
	\]
which is equivalent to:
	\[
	P(V \leq 0.5) = I_{1/2}(\alpha, \beta) < \frac{\beta}{\alpha + \beta}.
	\]

\noindent {\it  Proof} - 	Recall that:
	\[
	B(\alpha, \beta) = \int_0^1 t^{\alpha - 1}(1 - t)^{\beta - 1} \, dt,
	\]
and it satisfies the identity
	\[
	B(\alpha, \beta + 1) = \frac{\beta}{\alpha + \beta} B(\alpha, \beta).
	\]
Thus, the inequality $I_{1/2}(\alpha, \beta) < \frac{\beta}{\alpha + \beta}$ is equivalent to:
	\begin{equation}
		\int_0^{1/2} t^{\alpha - 1}(1 - t)^{\beta - 1} \, dt < \int_0^1 t^{\alpha - 1}(1 - t)^\beta \, dt. 
		\label{eq24}
	\end{equation}
Splitting the integral at $t = 1/2$ yields:
	\[
	\int_0^1 t^{\alpha - 1}(1 - t)^\beta \, dt = \int_0^{1/2} t^{\alpha - 1}(1 - t)^\beta \, dt + \int_{1/2}^1 t^{\alpha - 1}(1 - t)^\beta \, dt.
	\]
Substituting into (\ref{eq24}), we obtain:
	\begin{equation}
		\int_0^{1/2} t^{\alpha - 1}(1 - t)^{\beta - 1} \, dt - \int_0^{1/2} t^{\alpha - 1}(1 - t)^\beta \, dt < \int_{1/2}^1 t^{\alpha - 1}(1 - t)^\beta \, dt. 
		\label{eq25}
	\end{equation}
Simplifying the left-hand side:
	\[
	t^{\alpha - 1}(1 - t)^{\beta - 1} - t^{\alpha - 1}(1 - t)^\beta = t^\alpha (1 - t)^{\beta - 1},
	\]
so inequality (\ref{eq25}) becomes:
	\begin{equation}
		\int_0^{1/2} t^\alpha (1 - t)^{\beta - 1} \, dt < \int_{1/2}^1 t^{\alpha - 1}(1 - t)^\beta \, dt. 
		\label{eq26}
	\end{equation}
Now, perform the substitution $s = 1 - t$ on the right-hand side. When $t \in (1/2, 1)$, $s \in (0, 1/2)$, and $dt = -ds$, thus:
	\[
	\int_{1/2}^1 t^{\alpha - 1}(1 - t)^\beta \, dt = \int_0^{1/2} (1 - s)^{\alpha - 1} s^\beta \, ds.
	\]
Substituting into (\ref{eq26}), we obtain:
	\begin{equation}
		\int_0^{1/2} s^\alpha (1 - s)^{\beta - 1} \, ds < \int_0^{1/2} (1 - s)^{\alpha - 1} s^\beta \, ds. 
		\label{eq27}
	\end{equation}
Finally, for $s \in (0, 1/2)$, since $1 - s > s$ and $\alpha > \beta$, it follows that:
	\[
	(1 - s)^{\alpha - \beta} > s^{\alpha - \beta},
	\]
and multiplying both sides by $s^\beta (1 - s)^{\beta - 1}$ yields:
	\[
	(1 - s)^{\alpha - 1} s^\beta > s^\alpha (1 - s)^{\beta - 1}.
	\]
Thus, the integrand on the right-hand side of (\ref{eq27}) exceeds that on the left-hand side throughout $(0, 1/2)$, completing the proof.

\section{Concluding remarks}

We have presented an analytical framework for assessing the accuracy of collective decisions in networked populations, where agents interact through network-mediated dynamics and receive private signals biased toward the correct outcome. Classical models typically assume independent voters with individual accuracy exceeding chance, leading to near-certain correctness as population size grows. While some extensions account for weak correlations or shared external signals [e.g., 17, 18], our approach incorporates direct social influence, wherein each agent’s decision-making evolves dynamically based on the states of their network neighbors.

We show that, in the large-population limit, the probability that a majority selects the correct alternative converges to $1-I_{0.5} (\alpha,\beta)$, where $I_{0.5} (\alpha,\beta)$ is the regularized incomplete beta function. Although this probability generally remains below unity, it consistently exceeds the accuracy of any isolated agent whenever private signals are better than random—demonstrating that local imitation and social influence, typically associated with herding errors, can under suitable conditions improve collective performance.

This mechanism parallels cooperative behavior in physical systems, where weak external biases are amplified through local interactions to produce robust macroscopic order. In this sense, decision-making in social networks exhibits emergent dynamics reminiscent of interacting spin systems in statistical physics. Future work could explore how network topology, temporal dynamics, or heterogeneity in signal quality modulate this amplification effect, opening new avenues for understanding decentralized decision-making.

\begin{acknowledgments}
	M.A.M.A. would like to thank FAPESP, grant 2021/14335-0, and CNPq, grant 303814/2023-3 for financial support.
\end{acknowledgments}


\end{document}